\documentclass[
  ,final 
  ]
  {aipproc}

\usepackage{here}

\layoutstyle{6x9}

\begin{document}

\title{Summary of studies of the $\eta$ meson production with polarized proton beam at COSY-11}

\classification{14.40.-n, 14.40.Aq, 13.60.Le} 
\keywords      {$\eta$ meson production, production mechanism, analysing power}

\author{R.~Czy{\.z}ykiewicz~for~the~COSY-11~collaboration}{
  address={Institute of Physics, Jagellonian University, 30-059 Cracow, Poland}
}

\begin{abstract}
We report on the COSY-11 measurements of the analysing power 
for the $\vec{p}p\to pp\eta$ reaction and interpret the results in 
the framework of the meson exchange models.  

\end{abstract}

\maketitle

\section{Motivation}

In recent years the processes of the meson production have been 
extensively studied in the context of understanding the 
strong interaction, responsible for the existence of hadrons.  
The growth of the database of the observables connected with
the meson production in the hadronic collisions
made possible the verification of 
the predictions of the effective theories. 

A particular interest has been put on the studies of the 
properties of the $\eta$ meson~\cite{johanson, colin1}. Despite the fact that the discovery 
of this meson took place over forty years ago~\cite{pevsner} its production 
mechanism still remains unknown. Understanding 
of the production process of the $\eta$ meson may allow the theoretical 
models to be revisited with new input parameters: the coupling constants 
in the description of the production process of the $\eta$ meson, the initial 
and final state interactions and also dimensions of the reaction region. 

\begin{figure}[h]
\parbox{0.15\textwidth}{\hspace{0.8cm}
\includegraphics[width=0.2\textwidth]{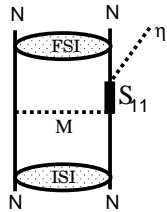}}
\vspace{-0.2cm}
\parbox{0.36\textwidth}{\caption{
The mechanism of the $\eta$ meson
production in nucleon-nucleon collisions.
M denotes an intermediate pseudoscalar or
vector meson, e.g. $\pi$, $\eta$, $\omega$, $\rho$.
ISI and FSI indicate initial and final state interaction
between the nucleons.
\label{pion}
}}
\end{figure}

From precise measurements of the total cross sections
of the $\eta$ meson production in the $pp\to pp\eta$
reaction~\cite{hibou,jureketa,bergdolt,chiavassa,calen1,calen2,moskal-prc,abdel}
it was concluded~\cite{moalem,bati,germond,laget,vetter,oset,nakayama,wilkin}
that this process proceeds through the
excitation of one of the protons to the S$_{11}$(1535)
state which subsequently deexcites via the emission of the $\eta$ meson
(see Fig.~\ref{pion}).
In practice, within the meson exchange
picture, the excitation of the intermediate resonance can be
induced by exchange of any of the
pseudoscalar or vector ground state mesons between the nucleons~\cite{bernard259,review,hab}. Based 
on the excitation function only it was, however, impossible  
to disentangle the contributions 
to the production process originating from the $\pi$, $\eta$, 
$\omega$ or $\rho$ meson exchange. \\
More constraints to theoretical 
models~\cite{moalem,bati,germond,laget,vetter,oset,nakayama,wilkin}
have been deduced from the measurement
of the isospin dependence of the total cross section
by the WASA/PROMICE collaboration~\cite{calen3}~\footnote{The
measurement of the isospin dependence is being extended by the COSY-11
collaboration~\cite{moskalpn,moskal0,joanna} to the $\eta^{\prime}$ production which may also be sensitive
to gluonic production mechanism~\cite{steven}.}.
The comparison of the $\eta$ meson production in
proton-proton and proton-neutron collisions
revealed that the $\eta$ meson
is by a factor of twelve more copiously produced when the total
isospin of the nucleons is zero with respect to the case when it 
equals to one. As a consequence an isovector meson exchange
is strongly favoured as being responsible for such a strong
isospin dependence.
This result was already a large step forward but still
the relative contributions of the $\rho$ and $\pi$ mesons
remained to be established.
For this purpose we have determined the  
analysing power for the $\vec{p}p\to pp\eta$ reaction since its 
theoretical value~\cite{nakayama,wilkin} 
is sensitive to the assumption on the type
of exchanged meson.

The first test measurement of the analysing power for the $\vec{p}p\to pp\eta$ reaction
at the excess energy of Q~=~40~MeV has been performed by the 
COSY-11 collaboration in the year 2001. The method of the 
analysis and the results have been reported in~\cite{winter1}.  
Unfortunately, the data from this tentative measurement are bared with
rather large error bars, and at the level of accuracy obtained in this 
experiment no constructive statement could have been done in order 
to distinguish between two different hypotheses of the $\eta$ meson production.
Similarly, the intepratation of the data obtained by the DISTO 
collaboration~\cite{disto}, performed in the far-from-threshold region 
at the excess energies of Q~=~324, 412, and 554~MeV suffered from the 
lack of a theoretical prediction for the analysing power. This is due 
to the fact that far from the reaction threshold the higher partial 
waves are involved in the reaction process, making the theoretical 
description complicated.

Further investigations were necessary and two additional 
experiments devoted to the determination of the analysing 
power for the $\vec{p}p\to pp\eta$ reaction in the close-to-threshold region have been performed 
by the COSY-11 collaboration.
In this paper we shall briefly present the experimental method 
and summarize results from these two measurements.
For the details of the analysis the interested reader is referred to the 
reference~\cite{czyzyk}.

\section{Method of measurement}

Experiments have been performed utilizing the COSY-11
facility~\cite{brauksiepe,klaja,smyrskikom} at the 
COoler SYnchrotron and storage ring COSY~\cite{maier}
in the Research Center J\"ulich, Germany. The analysing 
powers have been measured during two runs at different beam
momenta: p$_{beam}$~=~2.010~GeV/c (May 2003) and 2.085~GeV/c (September 2002), 
which for the $\vec{p}p\to pp\eta$ reaction correspond to the 
excess energies of Q~=~10 and 36~MeV, respectively. 

COSY-11 detection setup is described in~\cite{brauksiepe}.
A vertically polarised proton beam~\cite{stockhorst}
had been stored and accelerated in the COSY ring.
The target~\cite{domb} is installed in front of the accelerator's 
dipole magnet acting as a momentum separator for the 
charged reaction products. The positively charged ejectiles
are registered in drift chambers and scintillator hodoscopes. 
For each particle its direction of motion and time of flight on a nine
meter distance was measured. Tracking back these trajectories 
through the known magnetic field inside the dipole magnet to the 
reaction vertex allows for the momentum reconstruction.
Independent determination of momentum and velocity from the 
time-of-flight measurement permitted the identification of the charged particles. 
The $\eta$ meson has been identified
on the basis of a missing mass technique. 

For the determination of the analysing power of the $\eta$
meson at a given value of the polar angle $\theta_{\eta}$
and the azimuthal angle $\phi_{\eta}$ one has to measure a left-right
asymmetry of the yields of the $\eta$ meson production. 
The process of production is considered in the so called
{\it Madison coordinate frame}~\cite{madison}, which in our 
case has its $y$ axis parallel to the $\vec{p}_{beam}\times \vec{p}_{\eta}$
vector, with $\vec{p}_{beam}$ and $\vec{p}_{\eta}$ 
denoting the momentum vectors of the proton beam 
and the $\eta$ meson in the center-of-mass system, respectively. 
The $z$ axis of the Madison coordinate frame is along 
the $\vec{p}_{beam}$ vector, and the $x$ axis completes the 
right-handed coordinate frame. 

The COSY-11 detection setup is an asymmetrical aparatus, 
hence in the case of the $\vec{p}p\to pp\eta$ reaction, 
the acceptance for events where the $\eta$ meson is 
produced to the left side with respect to the polarisation 
plane is far larger as compared to the events where it is 
emitted to the right. Therefore, the left-right asymmetries
are determined from numbers of events with the $\eta$ 
meson production to the left side, measured for the spin 
up and spin down modes of the beam polarisation. 
Additionally, the acceptance of the COSY-11 facility 
allows to register only events scattered near the horizontal plane. 
In the analysis the azimuthal angle $\phi_{\eta}$ was restricted to 
values of $\cos{\phi_{\eta}}$ ranging between 0.87 and 1.

Let us define $N^{\uparrow}_{+}(\theta_{\eta})$ and
$N^{\downarrow}_{-}(\theta_{\eta})$  
as  production yields of the $\eta$ meson
emitted to the left around the $\theta_{\eta}$
angle as measured with the up and down beam polarisation, respectively, i.e.
\begin{eqnarray}
N^{\uparrow}_{+}(\theta_{\eta}) = \sigma_{0}\left(\theta_{\eta}\right)\left(1+P^{\uparrow}A_y\left(\theta_{\eta}\right)\right) E(\theta_{\eta}) \int{L^{\uparrow}dt} ,
\label{yields1}
\\
N^{\downarrow}_{-}(\theta_{\eta}) = \sigma_{0}\left(\theta_{\eta}\right)\left(1-P^{\downarrow}A_y\left(\theta_{\eta}\right)\right) E(\theta_{\eta}) \int{L^{\downarrow}dt},
\label{yields2}
\end{eqnarray}
with $\sigma_{0}\left(\theta_{\eta}\right)$ denoting the
cross section for the $\eta$ meson production for 
unpolarised beam, $P^{\uparrow\left(\downarrow\right)}$
standing for the polarisation degree corresponding to spin up and down modes,
E$(\theta_{\eta})$ being the efficiency of the
COSY-11 facility for detecting the $\eta$ meson
emitted to the left side at the $\theta_{\eta}$ angle
and $L^{\uparrow\left(\downarrow\right)}$ denoting the luminosity
during the beam polarisation up and down.
Signs in the brackets of Eqs.~\ref{yields1} and~\ref{yields2}
follow the Madison convention~\footnote{The detailed derivation 
of the Eqs.~\ref{yields1} and~\ref{yields2} can be found in~\cite{czyzyk}.}. 

Assuming that $P^{\uparrow}~\approx~P^{\downarrow}$~\footnote{Which is
valid within $\pm$2\% accuracy, as has been studied with the
EDDA~\cite{edda} and COSY~\cite{bauer} polarimeters.} 
and introducing the average beam polarisation
$P\approx\frac{P^{\uparrow}+P^{\downarrow}}{2}$,
the relative luminosity
$L_{rel}=\frac{\int{L^{\uparrow}dt}}{\int{L^{\downarrow}dt}}$
and solving Eqs.~\ref{yields1} and~\ref{yields2} for A$_{y}(\theta_{\eta})$
we obtain:
\begin{equation}
A_y(\theta_{\eta}) = \frac{1}{P} \frac{N^{\uparrow}_{+}(\theta_{\eta})-L_{rel}N^{\downarrow}_{-}(\theta_{\eta})}{N^{\uparrow}_{+}(\theta_{\eta})+L_{rel}N^{\downarrow}_{-}(\theta_{\eta})}.
\label{ayform}
\end{equation}

Therefore in order to calculate the analysing power 
one has to measure the relative luminosity $L_{rel}$, 
the average beam polarisation $P$, and the 
production yields $N^{\uparrow}_{+}(\theta_{\eta})$
and $N^{\downarrow}_{-}(\theta_{\eta})$.

\subsection{Relative luminosity}

The relative luminosity for both excess energies has been 
determined by means of the measurement of coincidence rate 
in the polarisation plane~\cite{czyzyk}. Due to parity 
invariance for strong interactions, the differential cross section for any two-body 
nuclear reaction in the polarisation plane does not depend on 
the magnitude of polarisation. Thus, the number of reactions
measured in the polarisation plane is proportional to the 
integrated luminosity over the time of measurement. 
The ratio of the numbers of events during spin up 
and down modes were used as a measure of the relative 
luminosity. Values of $L^{10}_{rel}=0.98468\pm 0.00056(stat)\pm 0.00985(sys)$
and $L^{36}_{rel}=0.98301\pm 0.00057(stat)\pm 0.00985(sys)$
have been obtained at the excess energies of Q~=~10 and 36~MeV, respectively.

\subsection{Polarisation}

The beam polarisation measurements have been performed with three independent detection
setups. In the run at the excess energy of Q~=~10~MeV the 
COSY-11 polarimeter has been used~\cite{czyzyk} as the main equipment 
to extract the information about the value of the polarisation degree. 
During this run a cross-check of the method was done by means of the COSY polarimeter~\cite{bauer},
and for the measurement at the excess energy of Q~=~36~MeV
the EDDA detection setup~\cite{edda} has been exploited.  

In all measurements the left-right asymmetry 
for the $\vec{p}p\to pp$ reaction was determined, and the polarisation was derived
using the data base of the analysing powers for the $\vec{p}p\to pp$ reaction, 
measured by the EDDA collaboration in the wide range 
of beam momenta and scattering angles~\cite{edda}.
For the detailed description of detector setups used for the polarisation 
determination the reader is referred to~\cite{czyzyk}. 
The values of polarisation degree $P^{10}=0.680\pm 0.007(stat) \pm 0.055(sys)$
and $P^{36}=0.663 \pm0.003(stat) \pm0.008(sys)$ have been obtained, for 
Q~=~10 and 36~MeV, respectively.

\subsection{Production rates}

\begin{figure}[h]
\includegraphics[width=5.0cm]{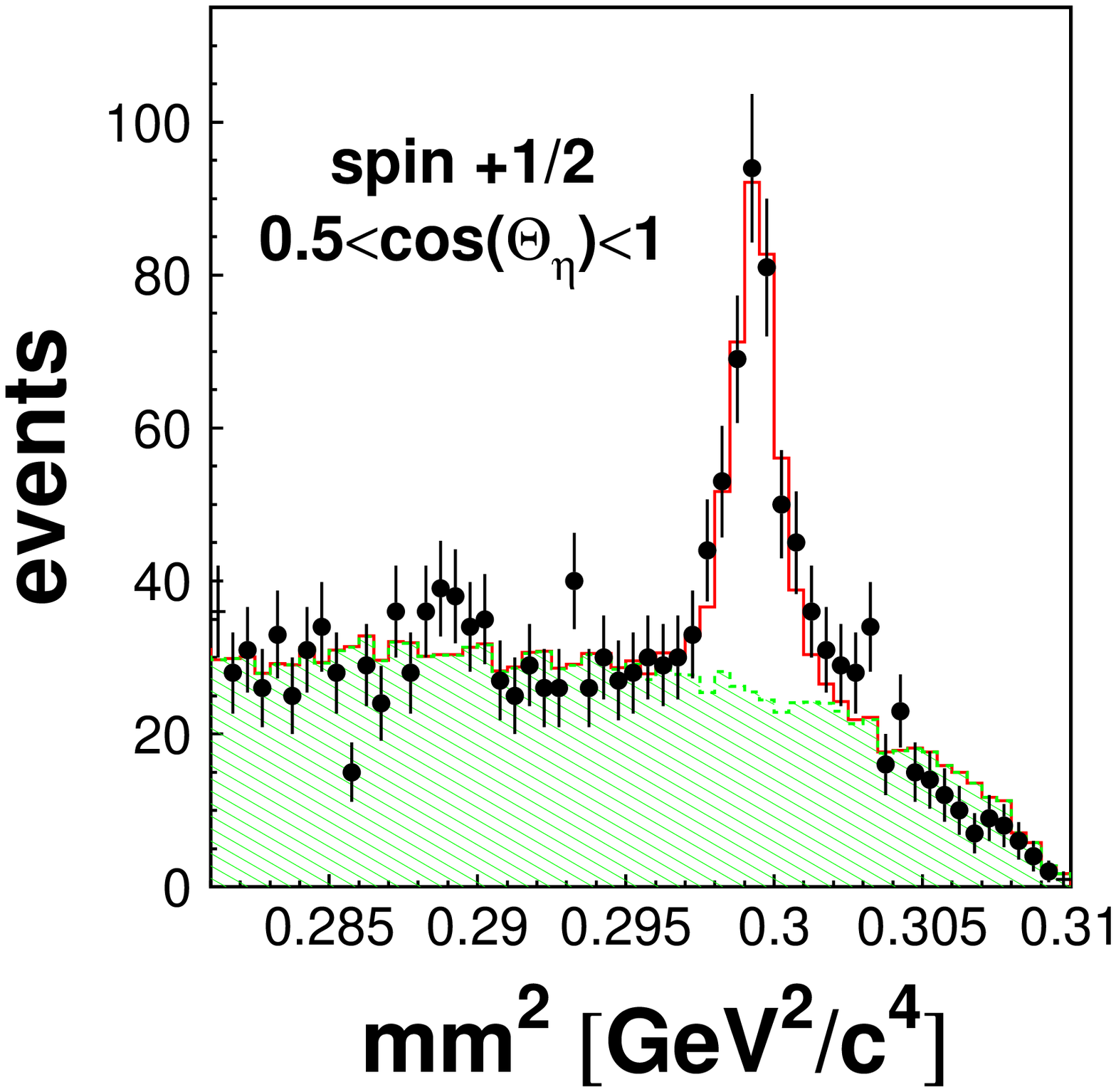}
\includegraphics[width=5.0cm]{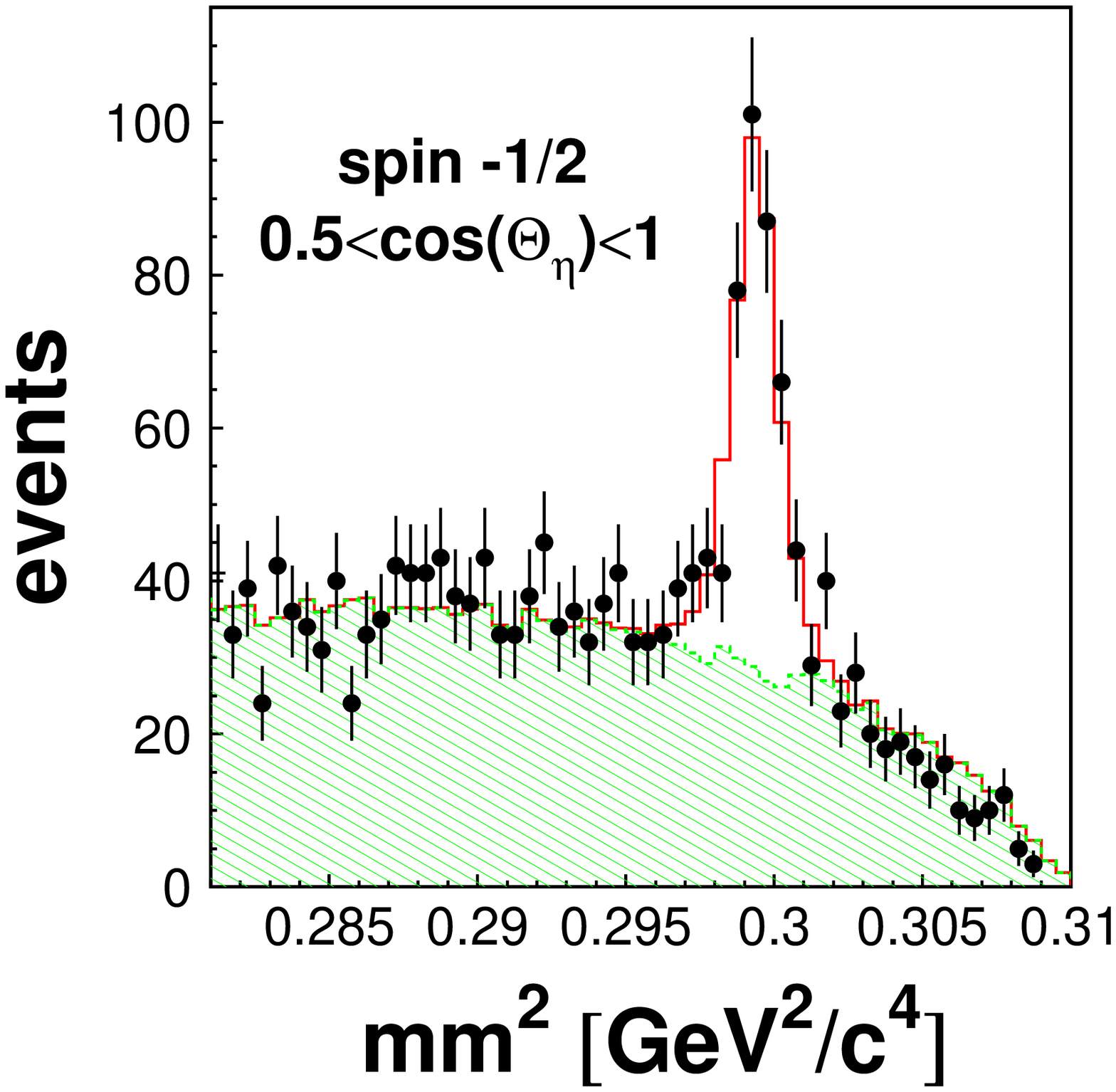}
\vspace{-0.2cm}
\caption{Examples of missing mass spectra for $\cos\theta_{\eta}\in\left(0.5;1\right)$
and opposite beam polarisation states, as measured 
at the excess energy Q~=~10~MeV. 
Full points correspond to the experimental values which are shown with their statistical 
uncertainties. The solid line represents the sum of the $pp\eta$ and multipion
background production channels determined by Monte-Carlo simulations.
The shaded parts of the histograms show the simulated 
contributions from  
the multipion background. \label{miss_mass}}
\end{figure}

The production rates $N^{\uparrow}_{+}(\theta_{\eta})$ and
$N^{\downarrow}_{-}(\theta_{\eta})$ from Eq.~\ref{ayform} 
have been extracted from the missing mass spectra. 
The range of the $\theta_{\eta}$ angle has been divided into 
four bins, at both excess energies~\cite{czyzyk}. 
The examplatory spectra of the missing mass distributions for the fourth bin 
as measured with spin up and down orientation at the excess energy 
of Q~=~10~MeV are presented in Fig.~\ref{miss_mass}. 

To separate the actual production rates from the background
both the reactions with multipion production as well as the events
with the $\eta$ meson production have been simulated using the
program based on the GEANT3 code. A fit of the simulated missing mass
spectra to the corresponding experimental histograms has been performed
with the amplitudes of the simulated spectra,
beam momentum smearing and the deviation of the beam momentum from its
nominal value treated as the free parameters. The integral of the Monte-Carlo spectrum
yielded the production rates. For more details on the
determination of the production rates
the reader is referred to~\cite{czyzyk}.

In the case of the measurement
at Q~=~36~MeV, the $\eta$ meson peak on the missing mass spectrum
was well separated from the kinematical limit and the
multipionic background could have been described by a polynomial of second order~\cite{czyzyk}.

\section{Results}

\begin{figure}[h]
\includegraphics[width=5.0cm]{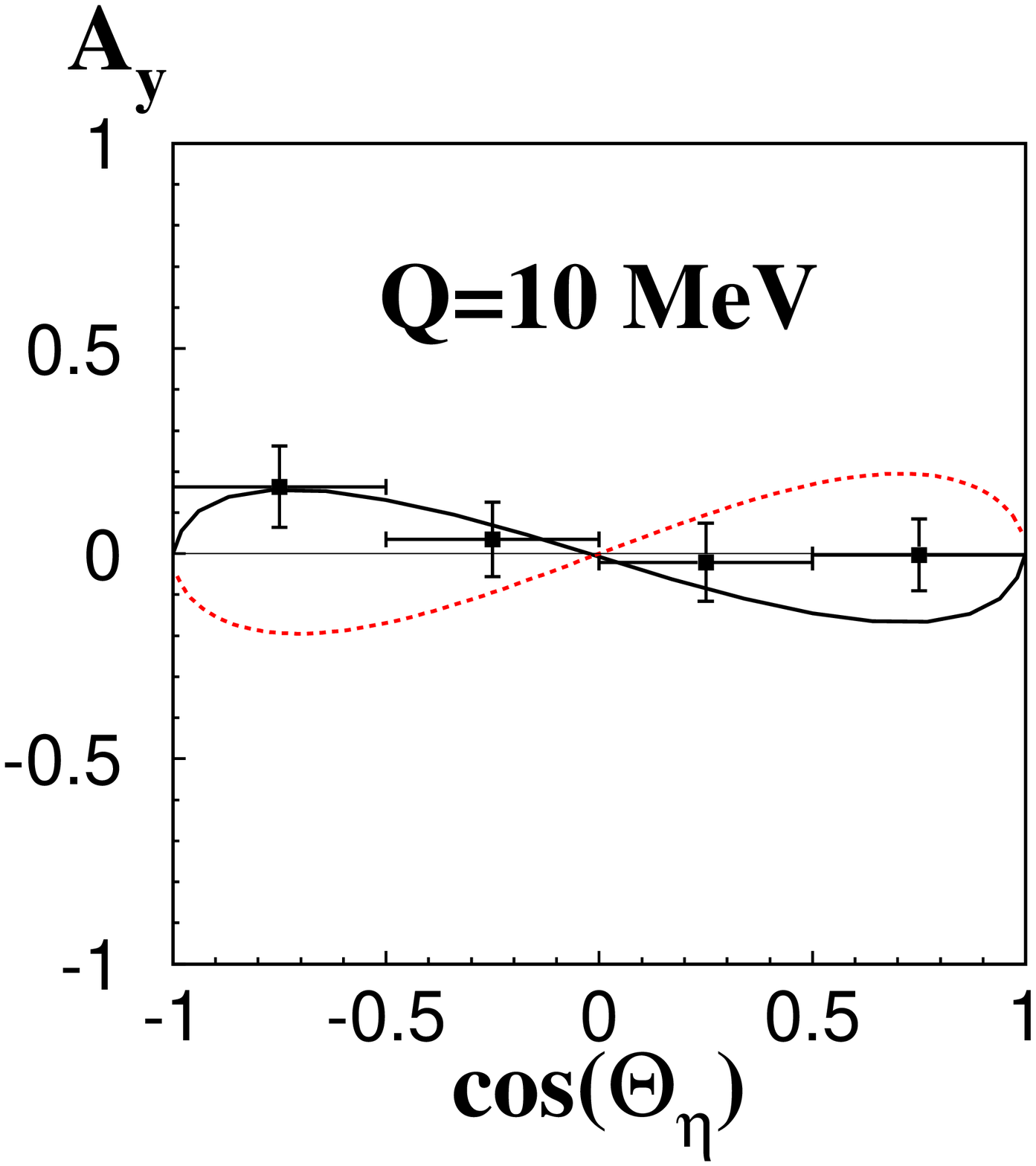}
\includegraphics[width=5.0cm]{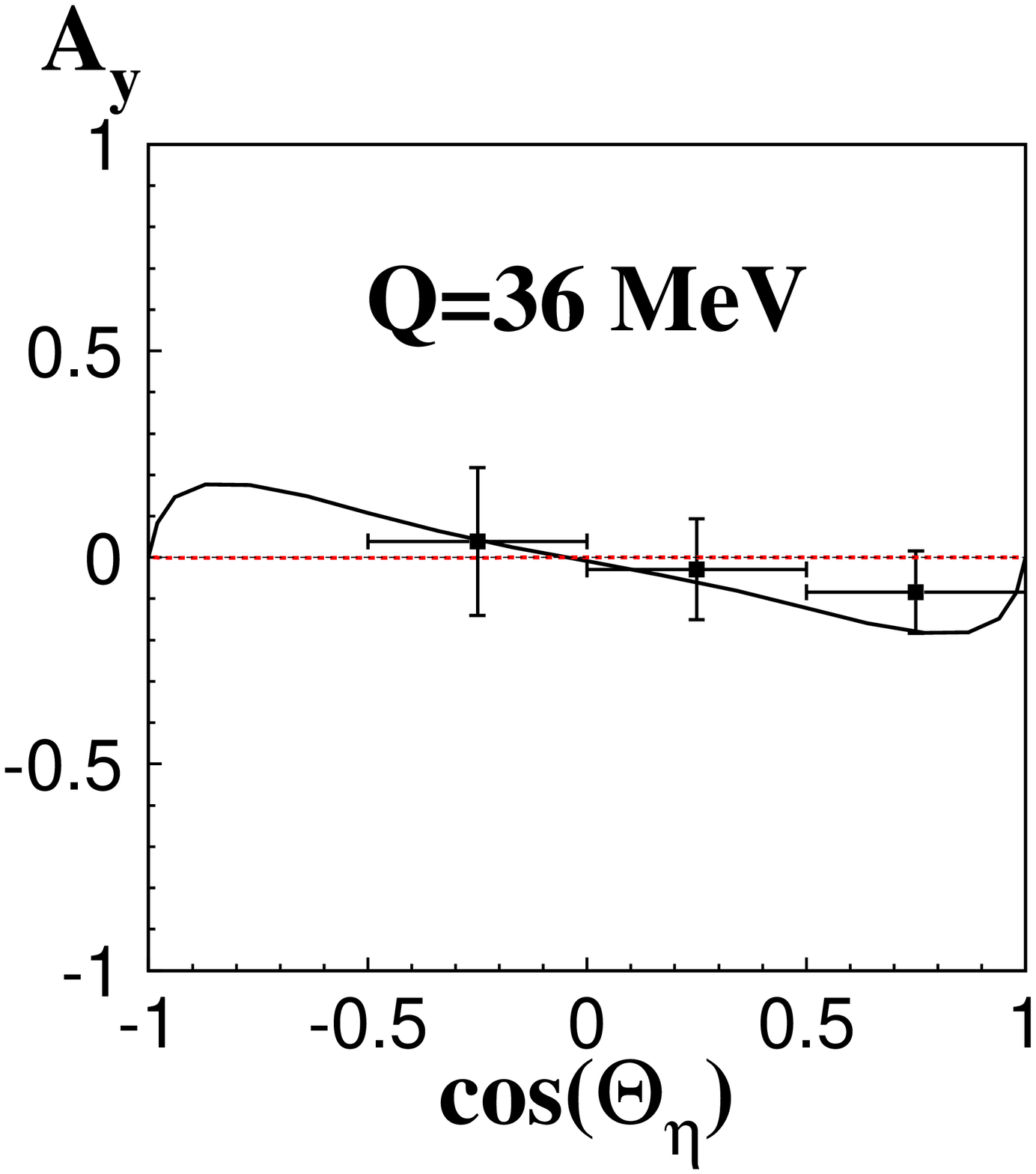}
\vspace{-0.2cm}
\caption{Analysing powers for the $\vec{p}p\to pp\eta$ reaction as functions of $\cos{\theta_{\eta}}$ for
        Q~=~10~MeV (left panel) and Q~=~36~MeV (right panel). Full lines are the
        predictions based on the pseudoscalar meson exchange model~\cite{nakayama}
        whereas the dotted lines represent the 
        calculations based on the vector meson exchange~\cite{wilkin}.
        In the right panel the dotted line is consistent with zero. 
        Shown are the statistical uncertainties solely. 
        \label{ay}}
\end{figure}

The analysing powers for both excess energies have been determined according to 
the Eq.~\ref{ayform} and are presented along with their
statistical errors in Fig.~\ref{ay}. The method of analysis is presented in details in~\cite{czyzyk}, 
and the results were already published in~\cite{czyzyk3,czyzyk33}. Tested predictions of reference~\cite{wilkin}
were based on the assumption of the $\rho$ meson exchange dominance and
the proton asymmetries taken from the photoproduction of the $\eta$ meson~\cite{photo}.
In the case of the calculations of reference~\cite{nakayama} the exchanges of 
all mesons have been taken into account in the framework of the relativistic meson 
exchange model of hadronic interactions and it was found in this model that the contribution from the 
pion exchange is  the dominant one.

The $\chi^2$ tests of the correctness of the models based on the dominance 
of the $\rho$~\cite{wilkin} and $\pi$~\cite{nakayama} meson exchanges have been performed. The reduced 
value of the $\chi^2$ for the pseudoscalar meson exchange model was determined to be 
$\chi^2_{psc}=0.54$, which corresponds to a 
significance level $\alpha_{psc}=0.81$, whereas 
for the vector meson exchange model $\chi^2_{vec}=2.76$, 
resulting in a significance level of $\alpha_{vec}=0.006$.

\begin{figure}[h]
\includegraphics[width=5.0cm]{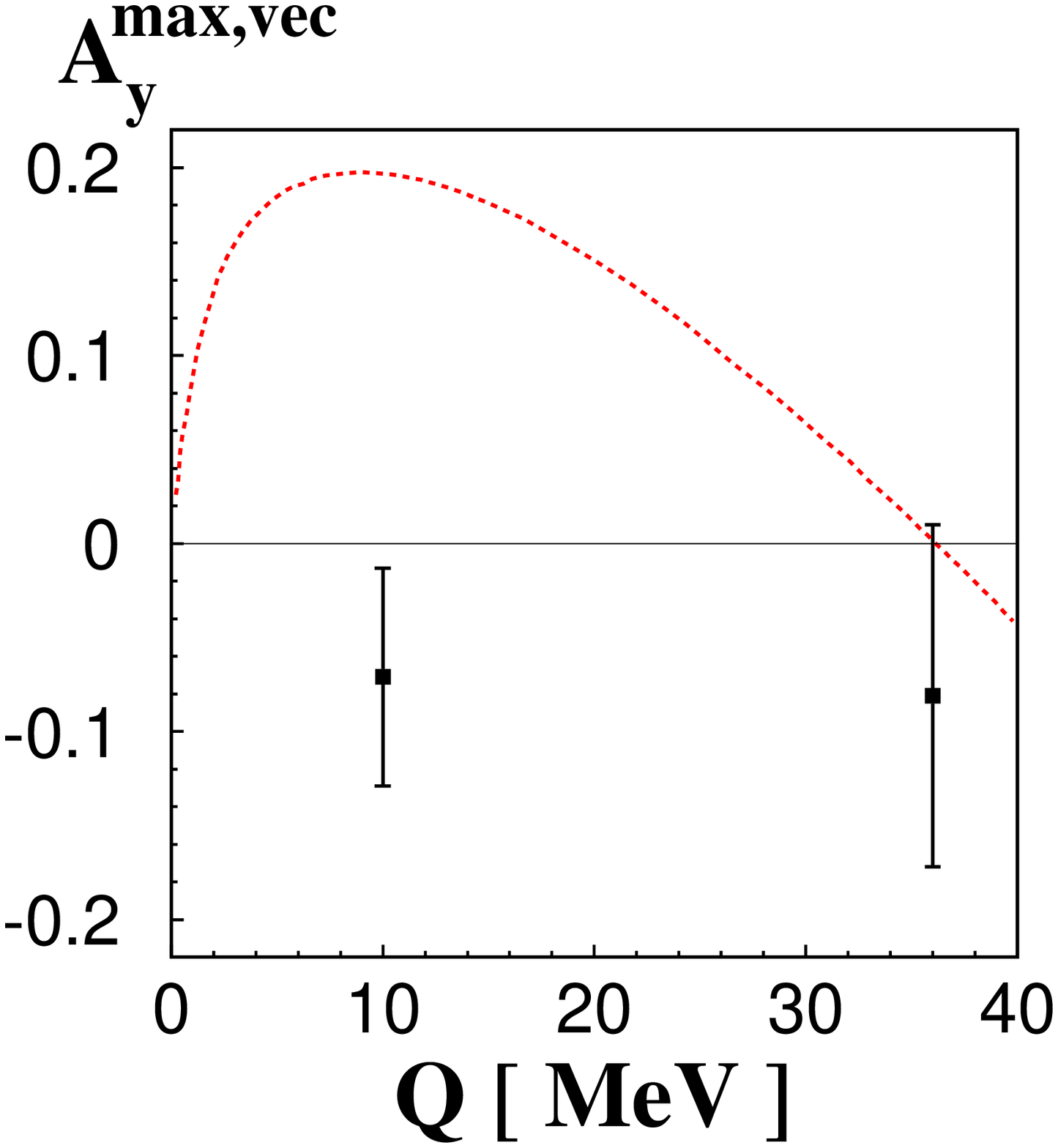}
\includegraphics[width=5.0cm]{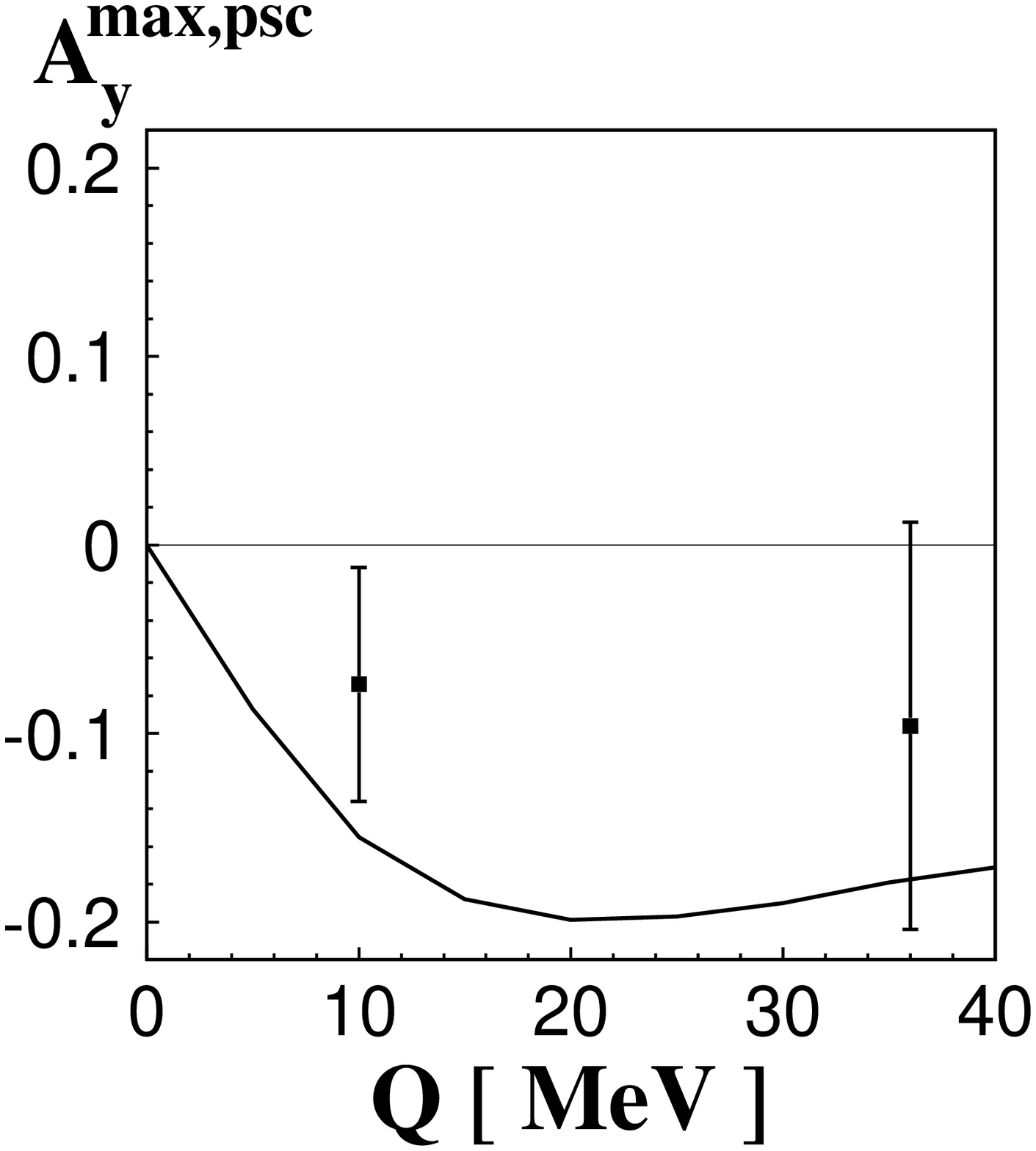}
\vspace{-0.2cm}
\caption{Theoretical predictions for the amplitudes of the analysing power's
        energy dependence 
        confronted with the
        amplitudes determined in the experiments at Q~=~10 and Q~=~36~MeV
        for the vector (left panel) and pseudoscalar (right panel) 
        meson exchange dominance model. \label{ay_max}}
\end{figure}

In the vector meson exchange dominance model~\cite{wilkin} 
the angular distribution of the analysing power
is parameterized with the following equation:
\begin{equation}
A_y(\theta_{\eta}) = A_y^{max,vec} \sin{2\theta_{\eta}},
\label{aymax}
\end{equation}
where the amplitude A$_y^{max,vec}$ is a function of the excess energy Q, 
shown as a dotted line in the left panel of Fig.~\ref{ay_max}.

We have estimated the values of A$_y^{max,vec}$ comparing the experimental data
with predicted shape utilizing a $\chi^2$ test. The values
of \mbox{A$_y^{max,vec}$ for Q~=~10} and 36~MeV have been determined 
to be \mbox{A$_y^{max,vec}(Q=10)~=~-0.071~\pm~$0.058} and  
\mbox{A$_y^{max,vec}(Q=36)~=~-0.081~\pm$~0.091}, respectively. 
Similar calculations have been performed for the pseudoscalar meson 
exchange model~\cite{nakayama}, assuming that the shape of the analysing power 
as a function of the $\cos\theta_{\eta}$ does not depend on the 
excess energy, which is correct within about 5\% accuracy.
It has been found that A$_y^{max,psc}(Q=10)~=~-0.074~\pm$~0.062,
and A$_y^{max,psc}(Q=36)~=~-0.096~\pm~$0.108.
These results are shown in Fig.~\ref{ay_max}.
The figure shows
that the predictions of the model based on the $\pi$
mesons dominance are fairly consistent
with the data, whereas
the calculations based on the dominance of the $\rho$ meson exchange
differ from the data by more than four standard
deviations. However, the latter calculation used the proton
asymmetry ($T$) in eta photoproduction~\cite{photo}, within the
framework of the vector meson dominance model~\cite{sakurai}, as the basis of
their estimate. It should be noted that it has proved hard to
reconcile the experimental value of $T$ with the results of
photoproduction amplitude analyses~\cite{MAID}.

\section{Conclusions and outlook}

Taking into account the $\chi^2$ analysis of the 
analysing power for the pseudoscalar and vector meson 
exchange models we have shown that the predictions of the 
pseudoscalar meson exchange dominance~\cite{nakayama} are in line with the 
experimental data at the significance level of 0.81.
On the other hand, the assumption that the $\eta$ meson is produced solely via the 
exchange of the $\rho$ meson~\cite{wilkin}, 
leads to the discrepancy between the theoretical predictions
and experimental data larger than four standard deviations. 
It must be stated, however, that the production amplitude for 
the $\rho$ meson exchange was determined based on the 
vector meson dominance hypothesis and the photoproduction data~\cite{photo}.
At this point it is also worth mentioning that the recent calculations of the 
$\eta$ meson production in the NN collisions performed in the 
framework of the effective Lagrangian model~\cite{shyam} also indicate 
the dominance of the pion exchange. 

The analysing power values for both excess energies are
consistent with zero within one standard deviation. This is 
in line with the results obtained by the DISTO~\cite{disto} 
collaboration in the far-from-threshold energy region.
Such a result may indicate that the $\eta$ meson is 
predominantly produced in the $s$-wave.

The improvement of the statistics would be possible  
with the measurements performed at the WASA-at-COSY facility~\cite{adam}. 
Thanks to instalation of a pellet target, high luminosities for 
the experiments with polarised proton beams are expected, 
promising to achieve the production yield of around
20000 $\eta$ mesons per day measured at the excess energy 
of Q~=~10~MeV. An experiment with such a high production 
rate within one weak would enable to reduce the error 
bars presented in Fig.~\ref{ay} by factor of 7. 
The letter of intent
for such an experiment has already 
been prepared by the COSY-11 collaboration.
For more details see~\cite{czyzyk2}.

\vspace{-0.4cm}
\begin{theacknowledgments}  
We acknowledge the support of the
European Community-Research Infrastructure Activity
under the FP6 programme (Hadron Physics, N4:EtaMesonNet,
RII3-CT-2004-506078), the support
of the Polish Ministry of Science and Higher Education under the grants
No. PB1060/P03/2004/26, 3240/H03/2006/31  and 1202/DFG/2007/03,
and  the support of the German Research Foundation (DFG) under the grant No. GZ: 436 POL 113/117/0-1.
\end{theacknowledgments}

\end{document}